*Perception! Immersion! Empowerment!*
# Superpowers as Inspiration for Visualization

Wesley Willett, Bon Adriel Aseniero, Sheelagh Carpendale, Pierre Dragicevic,
Yvonne Jansen, Lora Oehlberg, and Petra Isenberg

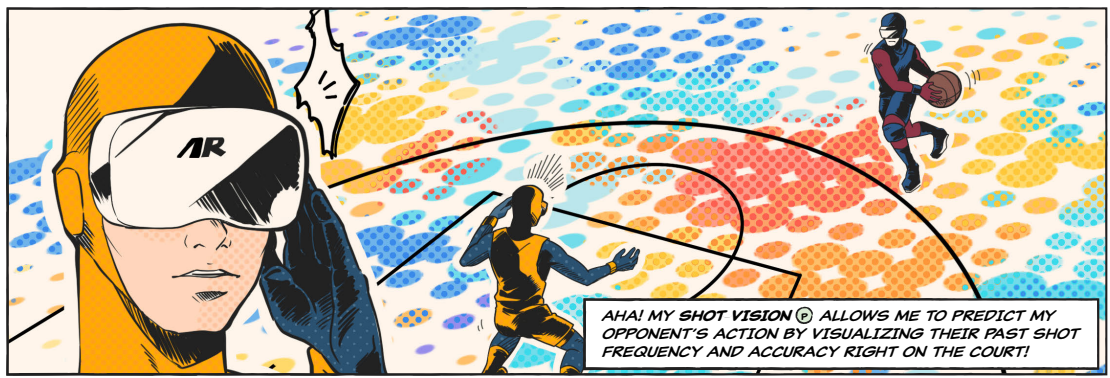

**Abstract**—We explore how the lens of fictional superpowers can help characterize how visualizations empower people and provide inspiration for new visualization systems. Researchers and practitioners often tout visualizations' ability to "make the invisible visible" and to "enhance cognitive abilities." Meanwhile superhero comics and other modern fiction often depict characters with similarly fantastic abilities that allow them to see and interpret the world in ways that transcend traditional human perception. We investigate the intersection of these domains, and show how the language of superpowers can be used to characterize existing visualization systems and suggest opportunities for new and empowering ones. We introduce two frameworks: The first characterizes seven underlying mechanisms that form the basis for a variety of visual superpowers portrayed in fiction. The second identifies seven ways in which visualization tools and interfaces can instill a sense of empowerment in the people who use them. Building on these observations, we illustrate a diverse set of "visualization superpowers" and highlight opportunities for the visualization community to create new systems and interactions that empower new experiences with data. Material and illustrations are available under CC-BY 4.0 at osf.io/8yhfz.

## 1 INTRODUCTION

When highlighting the benefits of visualization, many researchers refer to Card and colleagues' argument [18, 19] that visualizations "amplify cognition" in various ways, including expanding human working memory and enhancing recognition of patterns or outliers. Similarly, work on the "value of visualization" [88, 94] has focused on the knowledge-generating capabilities of visualizations that make people both more effective and efficient at understanding and analyzing data. Stasko [88] states that visualizations enhance humans' ability to gain insights from data and to understand the essence (or gist) of what the data contains, and to do so more quickly and with greater confidence. More rarely, visualizations are discussed as enhancing people's ability to engage with data intellectually, socially, physically, or emotionally [97].

The desire to enhance human perception, cognition, or experience is not unique to visualization. Fictional narratives in superhero comics, science fiction, and fantasy have long featured characters with extraordinary epistemic (knowledge-related) abilities that allow them to see, reason about, and understand phenomena that are otherwise invisible or incomprehensible. For example, Neo from the film *The Matrix* uses *accelerated perception* to track the movement of fast-moving objects like bullets. Destiny, an adversary of the X-Men, possesses *psionic precognition* (destiny perception), which allows her to see and predict future events. Characters such as these offer a rich and expansive sample of enhanced human abilities that can inspire new, impactful visualization techniques and systems—particularly as emerging technologies like extended reality, gestural interfaces, and ubiquitous sensing bring new opportunities for embodied interactions and situated visualizations.

We contribute two conceptual frameworks inspired by fictional superpowers that (1) characterize the common underlying visual, perceptual, and cognitive mechanisms that form the basis for many fictional abilities, and (2) describe ways in which visualization and other epistemic tools can give the people who use them a sense of objective or subjective *empowerment*—the increased ability to achieve goals. These frameworks are the result of three years of iterative discussions driven by our experience collecting, organizing, and scrutinizing depictions of enhanced epistemic abilities from a wide range of media including comics, film, television, and video games. During this process, we gathered a broad set of superpower examples, drawing on fan-curated superhero databases and our own exposure to superpowers in fiction. We then examined these abilities in the context of existing visualization and interaction technologies, as well as notions of empowerment from the human-computer interaction (HCI) and visualization literature.

The resulting conceptual frameworks showcase a variety of visual and cognitive mechanisms (including visual synesthesia, enhanced numeracy, and enhanced comparison), which can be used to critically

- *Wesley Willett and Lora Oehlberg are with the University of Calgary. E-mail: {wesley.willett,lora.oehlberg}@ucalgary.ca*
- *Bon Adriel Aseniero is with Autodesk. E-mail: bon.aseniero@autodesk.com*
- *Sheelagh Carpendale is with Simon Fraser Univ. E-mail: sheelagh@sfu.ca*
- *Pierre Dragicevic and Petra Isenberg are with Université Paris-Saclay, CNRS, Inria, LISN. E-mail: {pierre.dragicevic, petra.isenberg}@inria.fr*
- *Yvonne Jansen is with Sorbonne Université, CNRS, ISIR. E-mail: jansen@isir.upmc.fr*

examine existing visualization systems and to envision new ones. They also suggest concrete ways in which new systems can elicit a sense of empowerment by adapting how visualizations are displayed and controlled, as well as where and when they are shown. Inspired by these frameworks, we envision seven possible systems that highlight the potential for visualizations to augment human vision and cognition in ways that resemble fictional superpowers. Finally, we consider how a superpower-centric framing can impact discussions of the value of visualization, its usefulness as a source for visual design inspiration, dangers of this framing, and the importance of fairness and accessibility.

## 2 Related Work

The goal of much of human-centered computer science research is to help people extend their abilities—improving people's memory, motivation, organization, perception, etc. The work most closely related to ours investigates characterizations of human abilities and how one can enhance human abilities through system design.

### 2.1 Epistemic Activities vs. Other Human Activities

The HCI and cognitive science research literature has introduced terminology and constructs that distinguish between types of human activities, and by extension, human abilities. Several conceptual frameworks differentiate between activities related to knowledge acquisition and those that change the physical world. Kirsch [57], for example, distinguishes *epistemic* actions, which help people think and learn from the world, from *pragmatic* actions, which change the state of the world. Cadoz [16] draws a similar distinction between *epistemic* gestures and *ergotic* gestures, to which he adds the category of *semiotic* gestures—gestures used to communicate with other humans. Verplank [66] describes interaction with the physical world as three questions: How do you *feel* (perceive) the world? How do you *know* (cognitively understand) the world? How do you *do* (perform action) in the world? In our work we focus on epistemic actions and ways of knowing the world.

### 2.2 Enhancing Epistemic Abilities in Visualization

The quest to enhance or augment human capabilities is central to the visualization community. Munzner [69], for example, states that *"Vis systems are appropriate for use when your goal is to augment human capabilities [...]"*. Ware [98] even argues that *"people are not very intelligent without external cognitive tools"*. Visualization research focuses primarily on enhancing humans' epistemic abilities, specifically their ability to acquire knowledge through perception, cognition, and action. Card et al.'s list of enhanced human capabilities [19] highlights memory and pattern recognition. Ware [98] lists pattern recognition, perception of emergent properties, problem detection, multi-scale feature detection, and hypothesis generation as abilities that can be augmented by visualization. Spence [87] also discusses examples that demonstrate the huge time savings that visualization tools can bring to human data processing. Meanwhile, essentially all visualization (text)books emphasize visualizations' ability to trigger insights.

Discussions of the value of visualizations range from broad formulations that encompass both perception and cognition (such as Card [19] and Ware [98]) to narrower discussions of results from specific perceptual experiments. Individual studies often work from the bottom up, and measure the efficiency of specific data encodings on the perception of certain patterns (such as the length of lines) or for the enhancement of cognitive abilities (such as memory). Currently missing from the research discussions on the value of visualizations is a concrete characterization that summarizes benefits. In our work, we take a small step towards such a characterization to help researchers explore, discuss, and expand on the ways visualizations can impact people's ability to acquire knowledge and experience from data.

### 2.3 Fiction as a Critical Lens for Technology

To begin characterizing human abilities that are (or could be) augmented by visualization technologies, we look to fiction for inspiration. We use fiction as a means to better understand and convey the value of visualization, mirroring the growing use of design fiction and futuring approaches in related fields. User-experience designers and hardware developers regularly draw both implicitly and explicitly from science fiction [84] or use design futuring techniques to consider the social impact of new technologies [103]. Design fictions also offer an accessible entry point for complex topics, such as ethics in data science [68], sometimes in the guise of a mundane object like an IKEA catalog [15]. Multiple workshops have used fiction to inspire creative thinking in their participants [1, 17, 60]. HCI researchers have also begun to more explicitly engage with fiction—specifically science fiction—as inspiration for new research areas like shape-changing displays [93].

Engaging with fictional futures, technologies, or human abilities, can provoke designers and researchers and inspire them to reconsider the impacts of their work. Our aim is to help visualization designers reconsider how their contributions—visualization systems and techniques—might transform everyday people into heroes with enhanced epistemic abilities. We next discuss specific examples of interactive systems from research that are either inspired by fictional superpowers or are experienced as superpowers by the people using them.

### 2.4 Superpower-Inspired Interactive Systems

Many examples of sci-fi inspired interactive systems specifically seek to emulate the experience of superpowers with technology—especially in virtual reality (VR) and augmented reality (AR) research.

In some interactive systems, superpower metaphors are either explicit in the user interface or directly inspire its design. Ishibashi et al. [50], for example, proposed a VR system combined with a pulling force-feedback device that allowed players to "become" Spider-Man within their game, and haptically experience swinging through the air on elastic "webs." Rietzler et al. [81] proposed an airflow feedback system for VR and note that *"airflow simulation can go far beyond the simulation of wind, reaching from realistic effects to unrealistic superpowers for gaming."* Closer to our interest in epistemic-enhancing technology, Voss et al. [96] developed the *superpower glass* system based on Google Glass, which detects and labels others' emotions to help people on the autism spectrum develop emotional awareness. In an essay, Perlin [74] likens AR and wearable technology including hearing devices or Bluetooth hands-free cellphones to superpowers.

Some user studies have also been inspired by superpowers. For example, Imura et al. [70] explored how "hyper-natural" components of interaction techniques in VR may affect locomotion performance, while Rosenberg et al. [82] explored how the experience of superpowers like flight in VR may affect pro-social behavior.

Already, people using interactive systems often spontaneously liken them to superhuman abilities. For example, when Kajastila et al. [55] evaluated a game that taught motor skills using exaggerated visual feedback on a screen next to a trampoline, participants reported that the exaggerated feedback made them feel like they had superpowers. Bozgeyikli et al. [14], who compared several VR locomotion techniques, noted participants' *"very positive comments about the Point & Teleport technique resembling a superpower, being fun and being like a magical experience."* Similarly, in Raj and Ha-Brookshire's [78] interviews of professionals in the IT industry, many participants discussed wearable technologies as granting superpowers.

The idea that new user interface technologies can give people enhanced abilities is far from new. However, many of the previous systems and most prior discussions on the relationship between technologies and superpowers focus on *pragmatic* empowerment. Due to our focus on visualization, we address the challenge of facilitating knowledge-based, or *epistemic* empowerment through data.

## 3 Methodology

We first describe the process by which we arrived at our conceptual frameworks, then clarify and motivate their scope. A timeline and appendix documenting our process are included in supplemental material.

### 3.1 Process

We developed and refined our conceptualization of the relationship between visualization and enhanced abilities in fiction over the course of three years of discussions—including several in-person workshops and regular virtual meetings between members of the author team.

The initial concept emerged from a 2018 workshop on situated and embedded visualization. Inspired by the idea of "making the invisible visible" we discussed various reasons why data might be 'invisible', then generated a list of superheros whose abilities might render it visible. This triggered an ongoing series of post-workshop conversations about enhanced abilities and visualization, primarily grounded in well-known cinematic depictions of superheros, like Superman's *x-ray vision*, Spider-Man's *spider-sense*, and Iron Man's support from the artificial intelligence J.A.R.V.I.S. We expanded the list of abilities by immersing ourselves in online pop-culture encyclopedias, including the Fandom Superpower Wiki[1] (a community-curated database of superpowers that as of early 2021 included over 16,000 entries) as well as community-generated superpower taxonomies [21, 75]. We also revisited an extensive cross-section of original source material from movies and comics, focusing on specific characters whose abilities emphasize perception and knowledge generation. From this process, we distilled a list of 26 distinct perceptual and cognitive superhuman abilities from fiction, many of which are exhibited by a variety of characters. As we gathered each example, we looked at (a) which aspects of human abilities it amplifies, (b) in what contexts and situations characters use the ability, (c) how characters control the ability, and (d) how the ability is visually portrayed in print and other media.

After spending over a year exploring fictional superpowers and discussing 'empowerment' in visualization and HCI, we iteratively refined our list of epistemic superpowers, separating more complex superpowers into multiple distinct sub-powers. We also began to consider the technical and conceptual implications of each superpower in the context of visualization. This effort was led by one co-author, who presented drafts of the framework for "enhanced epistemic abilities in fiction" (described in Sect. 4) to fellow co-authors during weekly videoconferences, incorporating critique until the group reached consensus. The final framework highlights seven fundamental mechanisms that underpin the majority of the abilities found in our sample.

Based on this list of mechanisms, we then collected examples of real-world analogues to these abilities, including tools that operate on real environments (such as radiography) and in virtual environments (including both immersive and non-immersive visualization approaches). We performed this search using all seven mechanisms for enhanced abilities from Sect. 4 as well as five specific superpowers from our initial sample (*see-through vision*, *magnifying vision*, *night vision*, *shared vision*, and *mind reading*). Two co-authors led this process, presenting drafts and visual explorations in our weekly videoconferences to solicit additional examples, gather critique, and ultimately reach consensus.

Throughout our conversations, we increasingly reflected on the experience of using sensing and visualization tools and on how people's subjective and objective sense of empowerment from using these tools aligns with notions of superhuman empowerment. These discussions, grounded in our examples from the prior steps, ultimately motivated our seven *dimensions of empowerment* (Sect. 5), as well as our exploration of new possible abilities (Sect. 6), and higher-level reflections (Sect. 7).

---

[1] https://powerlisting.fandom.com/

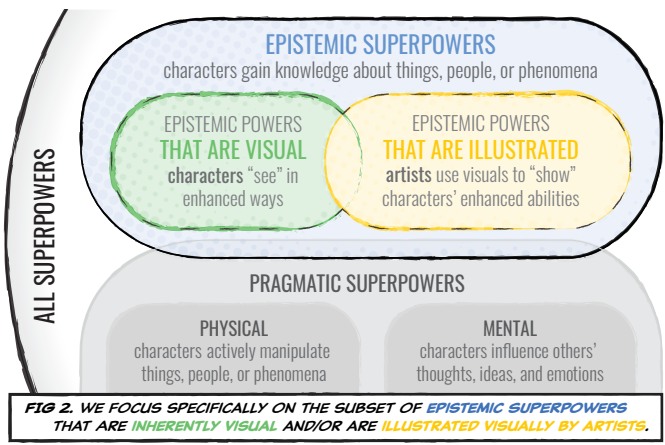

FIG 2. WE FOCUS SPECIFICALLY ON THE SUBSET OF EPISTEMIC SUPERPOWERS THAT ARE INHERENTLY VISUAL AND/OR ARE ILLUSTRATED VISUALLY BY ARTISTS.

## 3.2 Scope and Terminology

Based on our analysis of superpowers in fiction as well as prior work from HCI and cognitive science (see Sect. 2), we identified two broad classes of superpowers: *pragmatic superpowers* that enable characters to actively manipulate things, people, or phenomena in the world, and *epistemic superpowers* that allow characters to gain knowledge about the world without necessarily changing it (Fig. 2). In fiction, the vast majority of superpowers are pragmatic—including enhanced physical attributes (*super-strength*, *super-agility*, etc.) or manipulation abilities (*telekinesis*, *matter manipulation*, etc.).

In contrast, epistemic superpowers allow characters to gain knowledge of the world without necessarily altering it. These powers can take a variety of forms including literal extensions of traditional human vision (*x-ray vision*) and human cognition (*enhanced memory*), as well as extensions of other senses (*enhanced smell*). They also include more esoteric and indirect abilities (such as precognition of future events, perception of parallel dimensions, or the power to instantaneously compute probabilities) which allow characters to access and reason about information in ways not available to baseline humans. In this paper, we draw inspiration from the portrayal of *epistemic* empowerment in fiction to characterize the ways in which data and data visualizations can empower individuals. We are interested in how enhanced abilities for epistemic empowerment can offer people unique, site-specific, and context-specific assistance in the world.

When considering existing systems, we use the broad term *epistemic tool* to refer to any technological system that augments humans' ability to learn about the world. Examples include scientific instruments, optical devices, static information displays (such as maps, diagrams, or books), and computer-supported information systems, of which data visualization systems and visualization dashboards constitute a subset. Epistemic tools can extend human abilities to different degrees—ranging from trivial extensions to close-to-superhuman ones. In addition, we use the term *object(s)* to refer to the entities about which an individual learns when using an epistemic tool. In a data visualization context, objects are usually the physical referents (such as the physical spaces, objects, or entities) to which the data refers [102].

## 4 ENHANCED EPISTEMIC ABILITIES IN FICTION

Our first framework (see Fig. 3) attempts to capture low-level mechanisms that underpin a wide range of epistemic superpowers in fiction. We draw the connection to visualization by breaking down complex epistemic superpowers into a set of more atomic and composable abilities, often with analogues in perception and psychology research. These mechanisms include abilities that *enhance vision* (increasing humans' ability to use their visual system to observe phenomena in the surrounding world) as well as examples that *enhance cognition* (amplifying humans' capacity to process and reason about observations). We highlight seven specific classes of enhancements including *enhanced vision* (V) and *visual synesthesia* (Y), as well as enhanced *attention* (A), *numeracy* (N), *recall* (R), *comparison* (C), and *prediction* (P). These classes are not intended to be an exhaustive list but instead to provide a structure for examining enhanced epistemic abilities in fiction that suggest concrete opportunities for new visualization systems.

### 4.1 Enhancing Human Vision

Humans already rely on a variety of complex and interrelated mechanisms in our visual system (including color vision, depth perception, and stereoscopic vision) to collect information about our environments. The coordination and acuity of these underlying mechanisms differ considerably across individuals. Characteristics like color perception [51], visual search speed [95], and the size of the useful field of view [35] also vary with age or the presence of various genetic conditions. We use the blanket term *standard human vision* (Sv) to refer to the range of typical human perceptual experiences.

**Enhanced vision** (V) superpowers allow wielders to transcend limitations of the human visual system—granting characters the ability to see at extreme distances (the *telescopic vision* of DC's Superman), perceive very small objects (the *microscopic vision* of Marvel's Hyperion), or gain a wider field of view (the *360-degree vision* of Mad Eye Moody's magical eye from the Harry Potter series)—while still adhering to traditional human models of sight. Other characters can perceive electromagnetic wavelengths outside the standard human range. This can manifest in a variety of ways, including abilities like *night vision* (possessed by a wide range of characters including Marvel's Wolverine and Rebellion's Judge Dredd), which lets characters see in conditions with low or even no light. Other forms of enhanced vision include the ability to *see through occlusions and objects*, often selectively (Fig. 4). Some characters also possess the ability to see the world from viewpoints other than their own. Common ways of depicting enhanced vision include call-outs and close-ups, color-modified scenes, or establishing shots that provide larger fields of view.

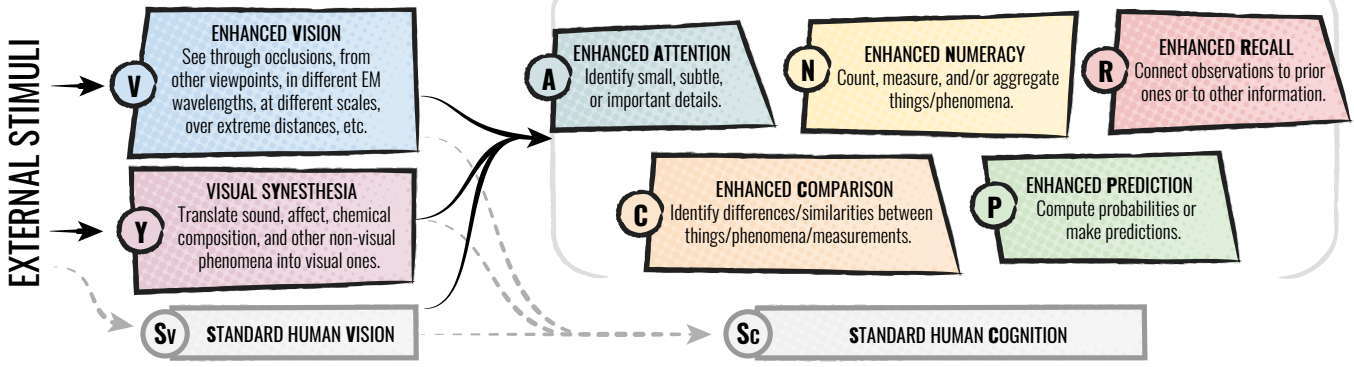

Fig 3. We characterize a set of underlying mechanisms behind fictional epistemic superpowers that extend characters' vision (left) and cognition (right).

Meanwhile, back in the real world, humans have long experimented with telescopes and related optical tools and, since the first x-ray image in the late 1800s, have developed a range of technologies for seeing *through* matter [67] using millimeter waves, terahertz waves, positrons, or ultrasonic waves [10, 56, 86, 92]. Some systems also allow people to effectively see through matter without directly sensing what is behind it. These approaches use object instrumentation or computer modeling plus AR to display occluded objects directly in the person's physical surroundings or on relevant objects [7, 40, 63, 76] (see Fig. 5).

Superpowers relying on **visual synesthesia** (Y), by contrast, allow characters to "see" non-visual properties of their surrounding environment. Typically, these abilities visually transform invisible phenomena like sound, affect, or chemical composition, allowing the wielder to perceive them in-place in the surrounding environment. Examples of visual synesthesia powers include the *emotion vision* used by DC's Black Lantern Corps (Fig. 4), the *chemical vision* of Marvel's Eye-Boy, or Wolf Link's *scent sight* in the Legend of Zelda. Because of the highly visual nature of superhero comics and related media, artists often illustrate variations of visual synesthesia (such as the *"sonar vision"* used by the blind Marvel character Daredevil) even in cases where the text or story specifically indicates some other perceptual mechanism.

Throughout human history, people have devised a range of epistemic tools to grant this ability—from thermoscopes and seismoscopes, which convert temperature and ground motion into visual form, to Chladni plates [33], which make the modes of vibration of rigid surfaces visible. More modern systems with sensory substitution [5] translate stimuli in the environment from one sensory modality to another—in fact virtually all information visualization systems translate data into a visual form [69]. Some systems (Fig. 6 [45, 100]) even overlay visualizations next to objects in physical environments [102].

### 4.2 Enhancing Human Cognition

In addition to augmented vision, numerous fictional characters possess enhanced cognitive abilities that extend their visual perception. These abilities allow them to process and reason about the world in ways that transcend *standard human cognition* (Sc).

Abilities that provide **enhanced attention** (A) allow characters to rapidly attend to important information or visual details in their environment that typical humans would otherwise miss. These powers are common for detective characters like Sherlock Holmes who seems unaffected by phenomena such as inattentional blindness, or Marvel's Spider-Man, whose *spider-sense* allows him to quickly direct attention to immediate threats. In media, attention-related abilities are often illustrated via visual highlighting (including outlines, glow, or explicit call-outs) and by using zooms and other re-framings of the scene to visually draw viewers' attention to specific details.

Humans already have the ability to quickly count small numbers of items [23] and make quick (though relatively imprecise) judgments about large numbers of items [91]. However, numerous fictional characters display **enhanced numeracy** (N) skills, which allow them to make rapid, confident, and accurate assessments about extremely large numbers of items. Often, these abilities extend to related types of quantitative perception, allowing characters to quickly and accurately judge distances, angles, weights, and other common physical measures. Daredevil, for example, can quickly count bullets even while distracted. Jason Bourne can estimate the exact weight of other people. Technologically-augmented, android, or (half) alien characters such

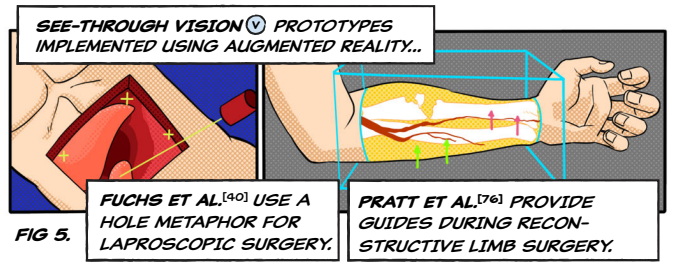

FIG 5. *See-through vision* (V) prototypes implemented using augmented reality... Fuchs et al.[40] use a hole metaphor for laproscopic surgery. Pratt et al.[76] provide guides during reconstructive limb surgery.

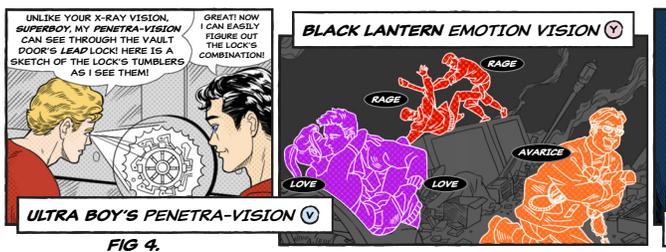

Ultra Boy's *Penetra-vision* (V). Black Lantern *Emotion Vision* (Y).

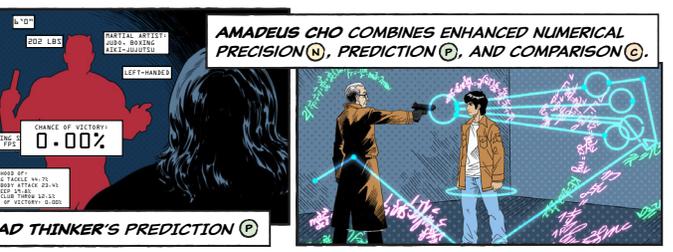

Mad Thinker's *Prediction* (P). Amadeus Cho combines enhanced numerical precision (N), prediction (P), and comparison (C).

FIG 4. Panels inspired by: *Superboy #98* (Jul 1962), *Blackest Night: Superman #3* (Oct 2009), *Daredevil: Road Warrior Infinite #3* (Mar 2014), *Incredible Hercules #137* (Oct 2009).

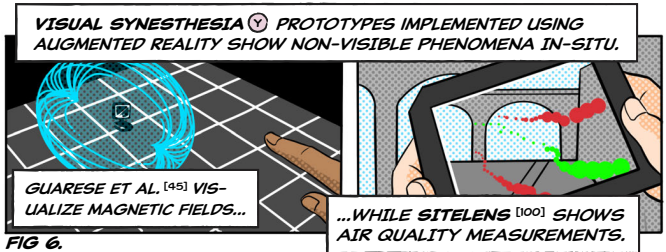

FIG 6. VISUAL SYNESTHESIA prototypes implemented using augmented reality show non-visible phenomena in-situ. Guarese et al. [45] visualize magnetic fields... while SiteLens [100] shows air quality measurements.

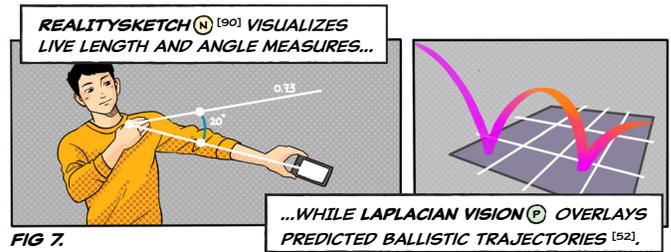

FIG 7. RealitySketch [90] visualizes live length and angle measures... while Laplacian Vision overlays predicted ballistic trajectories [52].

as *Star Trek*'s Data or Mr. Spock have shown the ability to quickly compute estimates (such as the number of tribbles). Most often artists have characters demonstrate their numeric abilities via dialogue and we found very few visual representations of the ability. Occasionally, special mathematical abilities are illustrated using floating numbers and equations in the scene or by highlighting the characters or objects that are being judged before the result is revealed.

Real-world epistemic tools that enhance numeracy, meanwhile, include counting devices (tally marks, mechanical counters, computer-vision-based counters [37]), calculating devices (abacuses, slide rules, computer spreadsheets), and technical measuring instruments (rulers, weighing scales, mass spectrometers). Visualization prototypes like RealitySketch (Fig. 7-left), also enhance numeracy by overlaying angular and distance measurements on real-world objects [90].

In other cases, characters may possess **enhanced recall**, which allows them to quickly and accurately recollect past observations from memory. These abilities help characters rapidly identify phenomena or connect new observations and information to prior experiences. Photographic (or eidetic) memory is particularly common in superhero fiction, shared by omnipotent characters like DC's Superman, technologically-augmented ones like RoboCop, and "peak human" characters like Ozymandias from Watchmen, Terry Sloane (DC's Golden Age Mister Terrific), or Jason Bourne. Characters' ability to access their memory or other related information is often illustrated using flashbacks and other visual devices that reveal past information to readers or viewers. In some cases, characters like Carrie Wells from the CBS crime drama *Unforgettable* are even shown examining and exploring their memory visually, walking into and examining past scenes. However, like numeracy, recall abilities are often simply described in dialogue, with characters referring to past information without illustration.

Similarly, characters can possess **enhanced comparison** skills, which allow them to quickly and accurately identify differences or similarities between phenomena. Comparison abilities are often related to and build upon characters' other skills like enhanced attention, numeracy, or recall. However, the ability to detect differences need not necessarily depend on quantitative judgments and may rely entirely on characters' other finely-tuned perception skills. For example the Marvel Universe Handbook [22] discusses how Daredevil *"can distinguish between identical twins at twenty feet by minute differences in smell."*

A variety of characters also exhibit the ability to reason about future events. In some cases these **enhanced prediction** abilities are described as true precognition, in which characters see glimpses of a single inevitable future. However, more often these abilities are shown as glimpses of possible futures—sometimes framed in probabilistic terms. In other cases (as with classic DC villain the Clock King or Marvel's Mad Thinker) these abilities are portrayed as purely logical or probabilistic, with characters making precise extrapolations based entirely on existing observations. Predictive powers are visualized in a variety of different ways across media, often by showing multiple possible outcomes either sequentially or in parallel or by accentuating the statistical presentation of outcomes (as in Fig. 4).

Real epistemic tools that support enhanced prediction include mechanical modeling and prediction systems (such as orreries, analog barometers, and analog ballistic computers [9]) as well as the multitude of predictive computer models in use today, from weather forecast models to route and trip planners. Recent AR prototypes like Itoh et al.'s Laplacian Vision (Fig. 7-right) have also demonstrated live predictions of object trajectories in real-world environments [52].

### 4.3 Composing and Combining Mechanisms

In practice, many empowered characters in fiction exhibit abilities that combine several of these underlying mechanisms. For instance, the detective Sherlock Holmes routinely displays attention, recall, and comparison skills that far surpass those of other individuals. In visual media, these abilities are often illustrated and integrated using variations on a visual trope (sometimes termed a *"Sherlock Scan"* [2]) that uses sequential zooms to highlight important or unseen details, flashbacks to provide context, and juxtapositions to reveal subtle differences that together facilitate the character's feats of deduction.

Similarly, Marvel superhero Amadeus Cho (in his various incarnations, including as the Totally Awesome Hulk) combines enhanced numerical precision, prediction, and comparison skills to make precise quantitative decisions (Fig. 4). In one common application, the character rapidly quantifies angles and trajectories in the space, then predicts and compares outcomes for each. Cho's ability is particularly notable from a visualization perspective, as artists have often chosen to illustrate it from the character's point of view—including numbers, trajectories, and other details that integrate into the surrounding scenes. As with Holmes, these abilities are usually explained as cognitive, with the characters still relying on typical human vision.

## 5 DIMENSIONS OF EMPOWERMENT IN EPISTEMIC TOOLS

As highlighted in the previous section, technological analogues to many superpowers already exist—allowing real people to see through walls, make precise numerical judgments, or predict future events. However, depending on *how* they are implemented, some existing systems create a much stronger sense of empowerment than others. Here, we examine real-world technologies (most from research areas other than visualization) that can provide some of the enhanced abilities that appear in fiction. We combine observations from these existing systems and their fictional counterparts, examining the varying degrees to which the people who use them *feel* empowered and more able to achieve their immediate goals. We highlight seven **dimensions of empowerment**: *scope*, *access*, *spatial relevance*, *temporal relevance*, *information richness*, *degree of control*, and *environmental reality*, exploring the manner by which each can alter people's sense of empowerment or agency. These dimensions go beyond the simple choice of visual mappings and spotlight visualization design considerations that are underexamined in current visualization research.

### 5.1 Scope

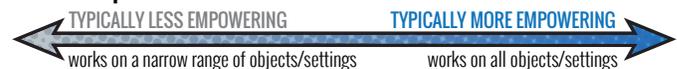

TYPICALLY LESS EMPOWERING ← → TYPICALLY MORE EMPOWERING
works on a narrow range of objects/settings — works on all objects/settings

The scope of an epistemic tool defines the set of objects or settings on which the tool can operate. Often, the broader the scope of an epistemic tool, the more empowering it is. For example, tourist binoculars mounted in a fixed location let people see visual detail (or, through AR, additional information [39]) from only one vantage point. In contrast, portable binoculars can be used in many settings, leading to a broader range of uses and greater agency. In the context of visualization, scope is often limited by the technical challenges of capturing and displaying data. For example, current electromagnetic sensing technologies that enable see-through vision (like x-ray and fMRI imaging) can typically only penetrate and detect specific types of materials and often involve complex immovable hardware. However, some exceptions offer greater portability, such as thermal imaging goggles, which can see through obstructions like smoke or fog [38].

---

[2]https://tvtropes.org/pmwiki/pmwiki.php/Main/SherlockScan

One way to broaden scope is to instrument environments with markers, sensors, transceivers, and other equipment that explicitly communicates information to a receiver. For example, Raskar et al.'s RFIG Lamps [79] allow people to see information about the content of warehouse boxes by having each box communicate its contents using active RFID tags. Another approach is to render a pre-existing computer model of the occluded objects. In one of the earliest AR concepts, a military aircraft pilot could see a 3D model of the landscape around them through the airplane hull and through clouds [11, 41]. Similarly, many mobile sky map apps include AR features that allow viewers to see the night sky during the day, through clouds, and even through the Earth itself (when pointing the phone down) [58]. Such tools do not require objects and settings to be instrumented but do require stable up-to-date models of them. As a result, the scope of current see-through vision systems based on object instrumentation and modeling is not larger (and is in fact often narrower) than sensing-based systems.

## 5.2 Access

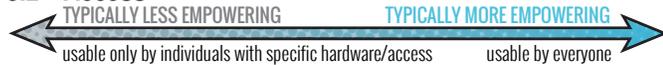

The access of an epistemic tool refers to who can use the tool and how easily. There are several ways that access can be facilitated or limited. Generally, tools that only exist at some specific locations (like a large particle accelerator in the Swiss Alps) have lower access than tools that can be easily copied, distributed, or used remotely. Access also relates to whether an epistemic tool can only benefit a single person or can be concurrently used by multiple people, and to how complex it is to equip or set up. For example, visual information provided by a head-mounted AR display is only accessible to the person wearing it, whereas anyone in the vicinity of a large screen or physical installation can access the displayed information. The relationship between access and empowerment is complicated but in general, the broader the access to an epistemic tool, the more empowering it is.

In the context of visualization, the increasing ubiquity of web-based systems and powerful mobile devices already highlights the potential for broad access. However, the creation of new and more empowering tools is still limited by the complexity and cost of sensing technologies, access to relevant data, and broader challenges related to visualization literacy [13]. Immersive and extended-reality technologies, which rely on hardware like head-mounted displays, also introduce new costs that can limit adoption, and are likely to create new asymmetries and coordination issues between people with access and those without [44].

## 5.3 Spatial Relevance

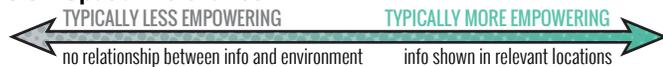

The spatial relevance of an epistemic tool reflects the distance (real or perceived) between the location where information is most useful to the person, and the location where that information is actually displayed. Since information about an object is often best interpreted in the context of the object itself, the most useful display location is frequently near or on top of the objects [102]. For example, a paper star map has lower spatial relevance than an AR app that overlays information directly onto the celestial dome. However, in some cases showing information next to the physical referent does not increase spatial relevance. For example, when someone seeks information about a distant referent (like a hotel to book), it is often preferable to show that information in front of the person (for example on a computer) rather than on the referent itself, where it might be invisible. In general, tools with greater spatial relevance are more empowering, allowing individuals to perceive and interpret information where it is more actionable.

Most contemporary sensor-based see-through displays, including tools like thermal cameras, already tend to have high spatial relevance. However, more complex scanning apparatuses like ultrasound imaging equipment typically still show information on a display separate from the object under inspection. AR research, meanwhile, has explored ways to overlay hidden objects directly on top of occluding ones [63], for example during laparoscopic surgery [40] or pregnancy ultrasound testing [7]. These research prototypes illustrate how high spatial relevance can facilitate decision making and action, and enhance the subjective impression of possessing super-powered see-through vision.

Similarly, most traditional measuring instruments are short-range and thus have moderately high spatial relevance, even when their displays are separate from their sensors. For example, most blood pressure monitors do not display blood pressure on top of someone's body, and a voltmeter does not display voltage readings on the electric circuit itself. However, by overlaying numerical displays on top of the physical environment, AR systems like RealitySketch [52] can increase spatial relevance dramatically. Using these kinds of AR approaches to overlay visualizations in real environments represents a clear opportunity for increasing spatial relevance for many other kinds of visualization applications, empowering people to translate observations into more immediate and relevant actions.

## 5.4 Temporal Relevance

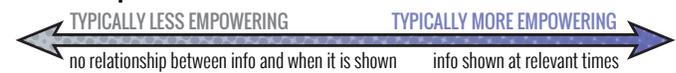

The temporal relevance of an epistemic tool reflects the temporal offset between the moment the information is delivered to the person, and the moment it would be the most useful for them to have the information. When it is important to react to current events, temporal relevance can often be achieved by showing real-time information [102]. For example, a live stock market display has a higher temporal relevance than the same information published in a daily newspaper. However, whenever information about the (possibly distant) past or future becomes useful to the person, high temporal relevance can be achieved by delivering this information at the present moment, even if the data is not directly connected to it. Often, higher temporal relevance increases empowerment, allowing individuals to more easily evaluate data and use that information to make immediate decisions.

The range of temporal relevance in current systems varies widely. For example, fast sensing-based see-through vision systems like airport luggage scanners have very high temporal relevance, showing imagery that is effectively live. In contrast, it currently takes up to an hour to produce fMRI images. Measuring instruments like thermometers can have low or high temporal relevance depending on how long it takes to calculate and display the numerical results. Most information visualization systems, meanwhile, tend to focus on visualizing data collected in the past. However, business dashboards and other decision-support tools are increasingly focused on visualizing real-time data [26].

Temporal relevance is particularly interesting for visualizations designed to enhance prediction, with advances in predictive modeling enabling predictions with increasingly finer temporal and spatial granularities. For example, while traditional television or newspaper weather forecasts focus on daily and regional predictions, new approaches can predict extremely local precipitation amounts on a minute-by-minute basis. Likewise, mobile traffic apps now provide near-real-time traffic predictions and visualizations, allowing drivers to dynamically adjust their routes based on predicted congestion and travel times. Prototype systems like Itoh et al.'s Laplacian Vision [52] or Alves et al.'s PoolLiveAid [2] go further, using AR and projection to visualize predicted object trajectories with particularly high spatial and temporal relevance. These approaches highlight the potential for many other visualization applications (including visualizing real-time predictions of auto and pedestrian behavior from autonomous vehicle models [101]) that could supercharge people's decision-making abilities.

## 5.5 Information Richness

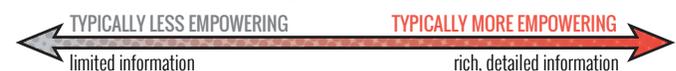

The information richness of an epistemic tool describes the quantity, variety, and accuracy of the information it is able to convey. In general, the higher the information richness of an epistemic tool, the more empowering it is.

Richness depends upon a variety of factors, including the quality and accuracy of the data as well as the fidelity of the output medium. For example, current real-world see-through vision approaches tend to be quite low-resolution and vary considerably based on the geometry and composition of the occluding objects. Tools for measuring and visualizing abstract and non-visible data, by contrast, already have the ability to convey much richer information. However, each instrument typically captures only one type of phenomenon: thermoscopes show temperature, seismoscopes show ground motion, etc. Although computer technology now enables data-agnostic methods for storing, transmitting, and visualizing data, data collection and sensing still require dedicated tools. Accuracy is also an important consideration, as a precise prediction can be thought of as carrying richer information than a vague prediction. While some epistemic tools make perfectly accurate predictions (such as predicting solar eclipses), others are very approximate (like long-term weather predictions).

### 5.6 Degree of Control

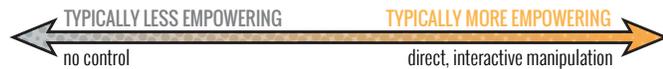

The degree of control of an epistemic tool refers to how much freedom the person has in activating and controlling the enhanced abilities that it facilitates. An always-on tool that cannot be tuned has very low control, while a configurable tool with many degrees of freedom that can be tailored to the situation offers more control. Degree of control also relates to *agency* (whether the person is controlling the tool [32, 53]) and *directness* (whether there is a lag or gap between the person's control actions and the outcomes [8, 53]). In general, the higher the degree of control of an epistemic tool, the more empowering it is.

In practice, many real-world sensors and visualization tools provide extremely limited control and expressivity. For example, medical or security-oriented scanners and scientific measurement tools like oscilloscopes typically have many settings, but these are rarely expressive or controlled via direct manipulation. Meanwhile, physical measurement implements like measuring tapes and protractors support extremely direct interaction, but have limited functionality. In contrast, prototype see-through vision systems like RFIG Lamps [79] can be directly aimed at objects. Other tools let people freely cut through virtual anatomical models by manipulating virtual [64] or physical [49] cutting planes. Similarly, prototype measurement and prediction tools (Fig. 7-left) like RealitySketch [90] and Laplacian Vision [52] have begun to explore opportunities to pair direct manipulation with automation in ways that empower people with a rich, direct, and powerful sense of control.

### 5.7 Environment Reality

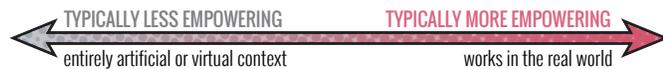

The environment reality of an epistemic tool reflects the extent to which its context is real rather than artificial. Tools that operate within the person's physical surroundings (as in Figs. 5–7) are higher along this dimension than tools that only work in specific virtual environments. Another important aspect of environment reality is whether the target objects are real or fictional. For example, a VR application that lets people explore real locations (say, a virtual visit to Paris) is higher on the reality spectrum than an application for exploring fictional worlds (like a virtual visit to Atlantis), irrespective of how realistic they are. Generally, tools that operate in real environments are more empowering, since decisions and actions there tend to be more consequential and lasting than those in virtual spaces.

However, creating complex and interactive epistemic tools is often much easier in virtual environments than in real ones. For example, see-through vision is easy to enable in virtual worlds, since the computer has a model of all occluded content. Many 3D video games allow players to see enemies through walls [4, 29] and—if the game is realistic and the immersion high—it is possible to evoke a subjective experience close to that of actually possessing the superpower. Related occlusion management techniques exist in many visualization systems, especially for 3D visualization [36]. These techniques are extremely powerful and allow people using computers to be much more effective at their tasks. However, they are rarely experienced as granting a superpower, in part because of their low environment reality.

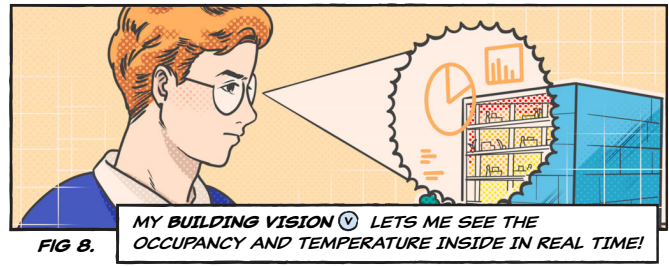

FIG 8. MY BUILDING VISION Ⓥ LETS ME SEE THE OCCUPANCY AND TEMPERATURE INSIDE IN REAL TIME!

Broadly, environment reality remains low for the vast majority of 2D and fully immersive [65] data visualization tools. However, data physicalization [54] approaches, particularly interactive and dynamic ones using technologies like drones [43] and robot swarms [59], highlight the potential for future visualization systems to empower people via increasingly real-world interfaces.

## 6 New Directions for "Empowering" Visualizations

Inspired by the understanding of epistemic superpowers and empowerment surfaced in our two frameworks, we propose a set of evocative new visualization designs and applications. While our designs are purposefully platform-agnostic and eschew details about their implementation, they highlight potential applications of the frameworks and underscore the breadth of opportunities for new, unconventional, and empowering visualizations that use situated and mixed-reality approaches [20, 85] to integrate into new settings and use cases.

**Enhanced Vision.** Ⓥ New visualization systems could build upon the metaphor of enhanced vision both literally and figuratively. For example, superimposing imagery from infrared cameras onto a viewer's field of view could enable *thermal vision*, making it possible to use heat signatures to detect thermal leaks, electrical faults, or fevers. In comparison to current on-screen thermal imaging approaches, techniques that directly augment the viewer's existing vision could widen the tools' scope while also increasing spatial and temporal relevance. By integrating thermal imagery into viewers' existing natural light vision, such approaches could also increase information richness, supporting new applications.

Other techniques could more dramatically alter a viewer's perspective, creating *panoptic vision* that allows them to see from new viewpoints. For example, a system might use cameras in an environment or on a viewer's body to build a 3D model of the surrounding spaces, then visualize it as a third-person world-in-miniature [25] showing the entire space, including their own body as well as areas outside of their view. Visualization designs can also borrow metaphors such as see-through vision for integrating sensed or simulated spatial data into real environments where occlusion is a problem—creating abilities like the *building vision* seen in Fig. 8.

**Visual Synesthesia.** Ⓨ Synesthetic visualizations that integrate data into viewers' real-world experiences can change how individuals approach tasks or appreciate the world around them. For example, using computer vision approaches to classify the emotional states of people in the immediate environment [83], then revealing them to viewers in-context could enable new kinds of *emotion vision* (Fig. 9). Like the emoji cloud visualization used in MeetCues [3], these tools could help a presenter track the changing emotional state of audiences during a meeting, and adapt presentations based on the audience's response, particularly benefiting individuals with neurodevelopmental conditions [28]. Such tools are likely most empowering when they operate in real-time, during an actual meeting, thereby offering high temporal relevance and high environment reality. Such approaches may also lend themselves to non-visual phenomena like sound, using audio processing tools to create dynamic *music vision* that visualizes musical tones, harmonies, rhythm, etc. around an audio source or a set of performers.

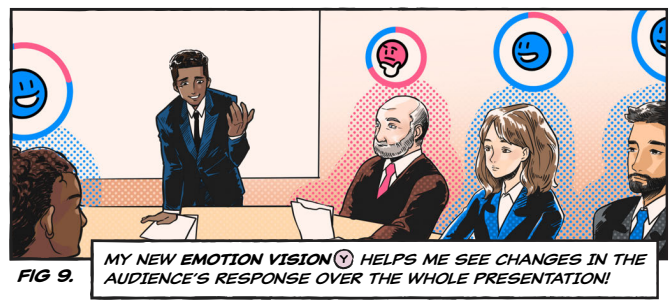

FIG 9. MY NEW EMOTION VISION Ⓨ HELPS ME SEE CHANGES IN THE AUDIENCE'S RESPONSE OVER THE WHOLE PRESENTATION!

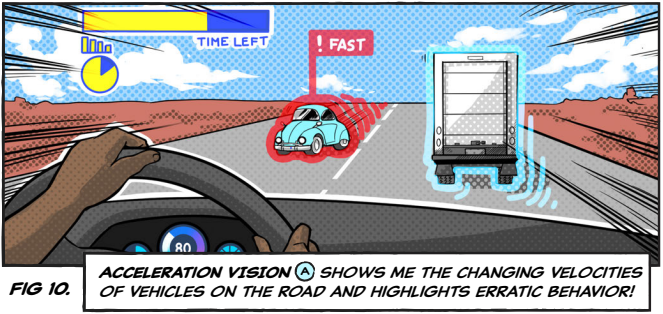

FIG 10. ACCELERATION VISION Ⓐ SHOWS ME THE CHANGING VELOCITIES OF VEHICLES ON THE ROAD AND HIGHLIGHTS ERRATIC BEHAVIOR!

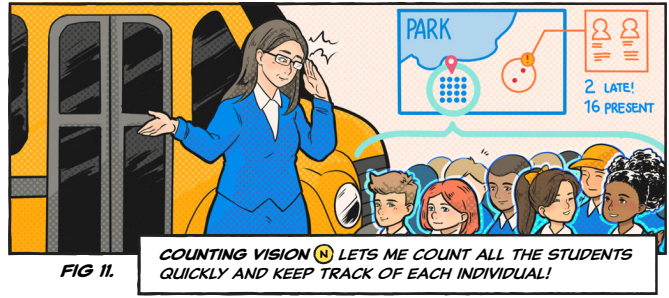

FIG 11. COUNTING VISION Ⓝ LETS ME COUNT ALL THE STUDENTS QUICKLY AND KEEP TRACK OF EACH INDIVIDUAL!

**Enhanced Attention.** Ⓐ Future visualization tools could also direct viewers' focus towards important objects or phenomena, helping filter out noise in both their data and environment. For example, an *acceleration vision* system integrated into a car's windscreen (Fig. 10) could visualize changes in acceleration of other vehicles, objects, or animals on a road, adding emphasis to draw attention to erratic or unexpected behavior. Similarly, *expression vision* might analyze facial micro-expressions for hundreds of faces at once, allowing an observer in a public space to evaluate public sentiment and highlighting faces meeting some predefined criteria, such as extreme joy or fear. Such attention-oriented approaches could extend highlighting techniques in existing visualization systems [80] or video games [30]. Alternatively, tools could also act implicitly to address viewers' attention biases by identifying when viewers disproportionately fixate on a subset of objects or values (one brand but not others, short/tall people but not average ones, etc.) then counteracting with *debiased vision* that highlights counterexamples or de-emphasizes overly-fixated ones. Such strategies could build on the growing appreciation of cognitive biases in visualization [31] as well as object subtraction [48], re-scaling [71], and distortion approaches for mixed reality use cases. Existing tools already empower people by enhancing their attention, but there are many opportunities to increase their subjective empowerment by seeking tools with higher levels on any of the dimensions of our empowerment framework.

**Enhanced Numeracy.** Ⓝ New tools for enhancing numeracy, meanwhile, could leverage computer vision to enhance people's ability to count, quantify, and analyze objects in their environment. *Counting vision* approaches could help individuals make precise and rapid judgements about large or abstract quantities like crowd sizes, numbers of specimens in a collection, or the volume of air in a building. These tools could also support common but high-consequence counting tasks like counting the members of a school class (Fig. 11). Meanwhile, more advanced *estimation vision* tools might help viewers more accurately calibrate estimations and judgements—providing in-context visual references that help compensate for judgements that people are characteristically bad at, including estimating volumes, comparing orders of magnitude, imagining exponential growth, or appreciating the uncertainty or variability in a set of measurements. In these scenarios, high temporal and spatial relevance is key to empowering people.

**Enhanced Prediction.** Ⓟ New and empowering predictive visualizations could extend enhanced numeracy approaches, allowing viewers to extrapolate from counts, trajectories, or other information. For example, a *queue vision* tool might allow individuals to instantaneously count the number of people in each queue at a supermarket checkout, then provide real-time predictions about which line will move fastest. Live predictive visualizations could also dramatically transform player behavior in team sports, taking the kinds of analytic visualizations that are now widely used by managers, coaches, and players off-court [73] and increasing the spatial and temporal relevance between these and the act of play. Visualization techniques like the *shot vision* imagined in Fig. 1 could display opponents' past shooting percentages, predictions of likely moves, and other information live—in an apotheosis of the increasing use of analytics in many leagues and sports.

**Enhanced Comparison.** Ⓒ Like prediction, new enhanced comparison approaches could build upon numeracy-oriented abilities, helping viewers more efficiently compare counts and quantities in a variety of settings. These might include powers like *measurement vision*, which extend the basic AR measurement tools already present on many consumer smartphones and letting viewers overlay scales, measurements, or previous observations onto new settings to support visual comparison. More advanced comparison techniques might allow viewers to reorganize physical space to support comparison using "remixed reality" methods like those explored by Lindlbauer and Wilson [61]. Visualization tools like these (Fig. 12) could empower people by allowing them to align, filter, cluster, and rearrange virtual copies of real-world objects to make their numbers, sizes, volumes, and other characteristics easier to compare—emulating the kinds of physical arrangements created by biologist Charles Davenport [27] or artists like Ursus Wehrli [99].

**Enhanced Recall.** Ⓡ Finally, techniques that enhance people's ability for recall have considerable potential in traditional visualization systems as well as for situated and immersive ones. For example, the visual language of flashbacks and "Sherlock Scan" revisitation of past observations could inspire new approaches to persistent challenges posed by visualizing interaction histories [47], analytic provenance [77], and collaborative analysis coverage [6]. In more immersive environments, new visualization tools could borrow even more literally from memory metaphors in fiction, creating systems that more explicitly emulate photographic memory. For example, immersive visualizations of extreme life-logging data—including the kinds of image, video, and metadata collections from systems like MyLifeBits [42] and challenges like ImageCLEF LifeLog [72]—could grant viewers new kinds of personal *history vision*, letting them visualize, summarize, and relive past experiences. AR information overlays also present opportunities for new visualizations that reveal complementary data to support in-place awareness and decision-making. For example, a *germ vision* system (Fig. 13) might use historical occupancy data to help cleaning staff prioritize key areas, while a *patient vision* system could provide doctors with AR summaries of past patient visits during consultations. Shared spatial history visualizations could also be extremely valuable and empowering for applications like search-and-rescue, allowing multiple searchers to see past search areas as well as gaps between them.

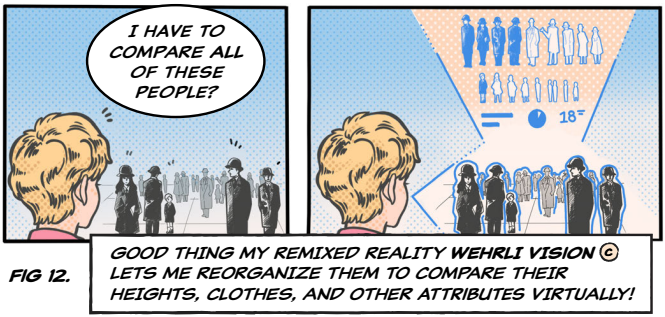

FIG 12. GOOD THING MY REMIXED REALITY WEHRLI VISION Ⓒ LETS ME REORGANIZE THEM TO COMPARE THEIR HEIGHTS, CLOTHES, AND OTHER ATTRIBUTES VIRTUALLY!

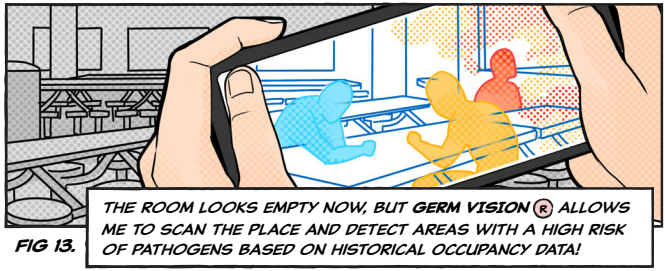

FIG 13. THE ROOM LOOKS EMPTY NOW, BUT GERM VISION® ALLOWS ME TO SCAN THE PLACE AND DETECT AREAS WITH A HIGH RISK OF PATHOGENS BASED ON HISTORICAL OCCUPANCY DATA!

## 7 DISCUSSION

In addition to these frameworks and new visualization concepts, our deep discussions over the last three years also probed several additional issues at the intersection of empowerment and visualization. In this section we highlight our most important threads of discussion.

**Superpower Depictions as Visual Design Inspiration.** While we highlight the potential for fictional superpowers to serve as inspiration for new kinds of visualization systems, our survey of visual superpowers in comics and other media also revealed a lack of consistency in how superpowers are represented and described. We anticipated that the visual representation of epistemic superpowers in these media, as well as the ways in which characters interact with and control them, might serve as direct sources of inspiration for new visualization designs. However, in practice, epistemic abilities in comics are often described rather than shown—typically as narration by the character using them. When illustrated, artists tend to treat visualizations as visual motifs or technobabble, signaling the complex nature of the abilities but providing little specific detail. As a result, we expect that most specific depictions of epistemic superpowers in fiction are unlikely to directly specify the design of visualizations. Nevertheless, the broad set of epistemic abilities imagined in these media suggest a diversity of new applications for visualization techniques (see Sect. 4).

**Visualization and Pragmatic Superpowers.** Our early discussions about these ideas narrowed the scope of superpowers to those most closely related to visualization—specifically epistemic powers that enhance how people "see" or which can be illustrated (Fig. 2). However, many visualizations can also be seen as giving their designers *pragmatic* powers over viewers. For example, narrative visualizations such as "The Fallen of WWII" [46] can evoke strong emotional responses in audiences. Other visualizations may be specifically designed for persuasion or to encourage behavior change (and, as a result, changes in the physical world). Considering these as pragmatic powers switches roles—empowering the designer rather than the viewer of the visualization—and reflects an opportunity for future exploration.

**Superpowers and the Value of Visualization.** The visualization literature already refers to a variety of human capabilities that visualizations can enhance [19]. However, when evaluating visualizations, researchers have tended to prioritize metrics of effectiveness and efficiency [94], despite the need to consider a wider range of ways in which new systems might provide value [97]. The lens of fictional superpowers could help designers and researchers broaden their view. Although superpowers like accelerated perception emphasize effectiveness and speed, fiction typically places a much greater emphasis on the value of the information itself. For example, superpowers that allow people to see the future, see behind surfaces, or see in the dark provide crucial and actionable information, regardless of how rapidly or easily this information is processed by the protagonist. Moreover, powers like the *ecological empathy* wielded by characters like DC's Poison Ivy or Marvel's Meggan Puceanu (who can sense and communicate with their environment) imply subtle affective and emotional benefits that are not well-captured by efficiency-oriented metrics.

Ultimately, the visualization community still lacks a concrete theoretical framework for characterizing the numerous ways in which visualizations can provide value. By highlighting the value of empowerment and its correlates, our *dimensions of empowerment* could serve as a starting point for a broader and more holistic framing, and also suggest new approaches for evaluating the benefits of visualizations.

A more encompassing framework could help communicate and justify the importance of new kinds of visualizations and application areas, encouraging new avenues for visualization research.

**Visualization Supervillains?** We (like most of visualization research) have chosen to focus on positive framings. However, the narrative dynamics of superhero comics—which typically pit super-powered heroes against villains who use similar abilities for more malicious ends—also lend themselves well to critical, adversarial, and "black hat" approaches to visualization [24]. In fact, even "positive" examples like our *emotion vision* or *building vision* could be easily reinterpreted as tools for surveillance and control. Meanwhile, approaches like the *shot vision* shown in Fig. 1 reflect a trend towards the increased use of analytics in that is already viewed with disdain by many sports fans, players, and journalists who assert that a focus on metrics and prediction undercuts the spontaneity of play. As such, the narrative structures of superhero fiction may serve as a useful tool for considering the negative and nuanced social implications of visualization systems, in addition to their potential benefits.

**Fairness and Accessibility.** In most fiction, superpowers are possessed by a very small fraction of people. When a superpower becomes widespread, it may cease to be considered a superpower (as Syndrome, the villain from *The Incredibles* puts it, *"When everyone is super, no one will be"*). Yet unlike superhero fiction, which largely focuses on confrontation and conflict, visualization research tends to envision futures in which increasing numbers of people have access to epistemic tools and society as a whole becomes more empowered.

However, considering societal empowerment also requires a deeper consideration of equity, fairness, and accessibility. Currently access to epistemic tools varies widely due to social and economic factors, as well as the immense variability of human visual and cognitive abilities [62]. Visualization research has already made strides in addressing specific disabilities like color vision deficiency and a few projects have focused on empowering individuals with more profound vision impairments [34, pp.11–13]. Looking forward, new epistemic tools might even help individuals with visual or cognitive differences gain abilities (à la Daredevil) that exceed those of typical humans. More broadly, understanding the potential social impact of epistemic tools calls for models (perhaps improving upon van Wijk's [94]) that consider the value of increasing overall human empowerment *and* equity.

## 8 CONCLUSION

In this paper we have explored the potential for superpowers to serve as a source of inspiration for visualization, examining visualization-related powers in fiction, considering their real-world analogues, and imagining new visualization superpowers of our own. We intend this work as a provocation and an encouragement to the visualization community to actively examine more diverse sources of inspiration for visualization designs, applications, and theory. While our exploration draws predominantly on western superhero comics, many other media also explore possible augmentations of human abilities in ways that may prove inspiring to visualization designers and researchers. Adjacent domains like manga and anime, as well as science fiction and fantasy more broadly, are brimming with visual references, metaphors, and other ideas relevant to visualization. Exploring parallel genres like these presents opportunities for visualization research to capitalize on the kinds of design futuring and speculative design that are increasingly being explored in HCI [12, 84, 89]. We expect that these approaches can encourage creative and divergent thinking about the future of the field, especially as new platforms and use cases make visualizations an increasingly ubiquitous part of our lives.

### ACKNOWLEDGMENTS

We thank all of the SEVEN workshop participants and colleagues who contributed to the genesis of this work. This research was funded in part by Inria-Calgary associated team SEVEN, the Canada Research Chairs Program, the Natural Sciences and Engineering Research Council of Canada (NSERC), Alberta Innovates Technology Futures (AITF), SMART Technologies, ULC, and the ANR grant ANR-19-CE33-0012.